# MetaSort: An Accelerated Approach for Non-uniform Compression and Few-shot Classification of Neural Spike Waveforms

Luca M. Meyer and Majid Zamani, *Member*, *IEEE*

*Abstract*— **Many previous works in spike sorting study spike classification and compression independently. In this paper, a novel algorithm is proposed called MetaSort to address these two problems. To deal with compression, a novel adaptive level crossing algorithm is proposed to approximate spike shapes with high fidelity. Meanwhile, the latent feature representation is used to handle the classification problem. Besides, to guarantee MetaSort is robust and discriminative, the geometric information of data is exploited simultaneously in the proposed framework by meta-transfer learning. Empirical experiments with in-vivo spike data demonstrate that MetaSort delivers promising performance, highlighting its potential and motivating continued development toward an ultra-low-power, on-chip implementation.**

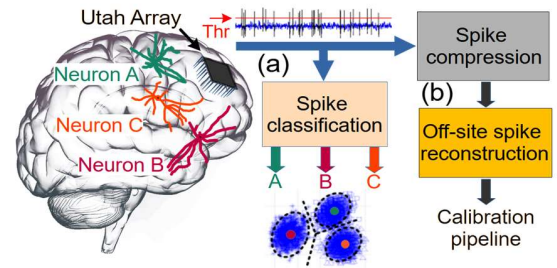

Fig. 1. (a) Spike classification for determining single unit activity. A, B, and C are the classified neurons. (b) Spike compression head that performs almost loss-less approximation on spike waveforms for off-site re-calibration or re-sorting in cases of low on-chip sorting confidence.

## I. INTRODUCTION

Extracellular recordings have been widely used to monitor neuronal activity by implanting multi-electrodes in the cortex and capturing high-dimensional neural data. A processing step, known as spike sorting, is necessary to separate threshold-crossing multi-unit activities and assign the captured spikes to their originating neurons [1]-[4]. Spike sorting is an invaluable signal-processing tool applied in brain-machine interface (BMI) research for studying and decoding neural signals from different brain regions and understanding the underlying mechanisms [5].

A recent trend in brain recording is the utilization of high-channel count neural interfaces that include tens of thousands of sensing sites [6]. Large-scale data streaming to a remote computer for processing is possible, however extremely challenging due to the data size (i.e. Gbps) and limits the viability of experimental setups. This necessitates efficient on-implant data compression due to restrictions in both the wireless transmission bandwidth and the power budget allocated for the telemetry of the recorded neuronal activities.

Numerous temporal and spatial compression approaches have been developed to reduce the transmission bandwidth while maintaining the waveforms of individual spikes. Temporal compression techniques include discrete wavelet transform (DWT) [7], Walsh-Hadamard Transform (WHT) [8] and Taylor series [9]. Conversely, spatial compression methods examine the patterns seen in spiking activities. Among the successful spatial compression techniques reported in the literature are compressive sensing [10] and the deep autoencoder [11].

This paper introduces a new approach called MetaSort, that combines spike compression and classification as shown in Fig.1. MetaSort employs a derivative-informed level-crossing sampling strategy to retain the most informative points of each spike while discarding redundant data. A multi-task artificial neural network (ANN) jointly performs compression index prediction and spike classification, while a lightweight few-shot learning module enables rapid adaptation across recording channels with minimal labelled data. Validation on high-density extracellular recordings demonstrates that MetaSort achieves classification accuracy above 94.4%, 6X compression with averaged 0.05 root mean squared error (RMSE) tested on 500 spikes, and few-shot adaptability across diverse conditions.

## II. METASORT CORE DESIGN

Fig. 2 shows MetaSort in training, mapping and calibration phases. During the training phase, an adaptive level crossing sampler identifies informative temporal samples of the spike waveforms for a multi-task ANN with a compression head that predicts the positions of retained sampling points. The classification head outputs the probability distribution over the neuron and artefact classes. Because the two tasks are trained jointly, the encoder learns to extract features that are simultaneously compact and discriminative, improving efficiency without sacrificing accuracy. Meta-transfer learning (MTL) is embedded onto the calibration phase which provides robustness under distributional shifts across recording channels. This selective update strategy reduces computational cost while maintaining stable performance, enabling rapid recalibration in practical neural recording scenarios. In the last layer, the ANN splits into two different fully connected layers. The main branch returns the preserved spike samples $(S_1, S_2, ..., S_8)$, while the auxiliary branch assigns class labels to spikes and artefacts (i.e. S/A) as shown in Fig. 2(a). Once the

L. M. Meyer and M. Zamani are with the School of Electronics and Computer Science (ECS), University of Southampton, Southampton SO17 1BJ, U.K. (e-mail: L.M.Meyer@soton.ac.uk, M.Zamani@soton.ac.uk).

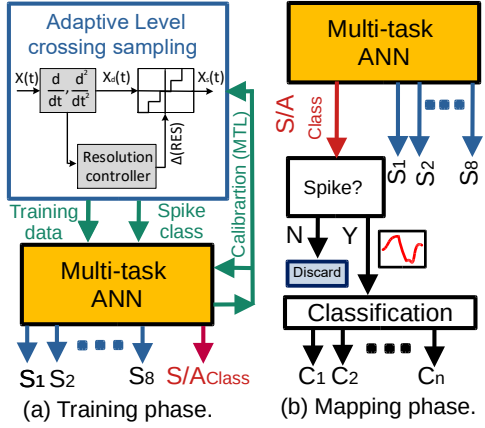

Fig. 2. MetaSort in (a) learning phase and (b) optimal feature extraction. The core of the MetaSort comprises components including an adaptive level crossing sampler, a multi-task ANN, and a MTL unit.

training phase is performed, the second phase, cluster mapping, is initiated. At this stage, the spikes from the recording channels are simultaneously compressed ($S_1, S_2, \dots, S_8$) and assigned to one of the cluster means identified in the training phase ($C_1, C_2, \dots, C_n$) as shown in Fig. 2(b). The recorded action potential shapes may vary during the cluster-mapping phase. Fast adaptation is triggered in the ANN and sampling units using MTL to update the network parameters on the recording channel if there is mismatch between the input spike and the class to which it is assigned.

### A. Pre-processing

Neural waveforms were synthesized with a spike generator [1], [2] displayed in Fig. 3(a). A standardized pre-processing pipeline is adopted that aligns and normalizes spike waveforms before they are processed by the adaptive level crossing sampling unit. Peak-based temporal alignment is applied to remove temporal jitter. For each extracted waveform $x = \{x_i\}_{i=1}^{64}$, the maximum peak $p = arg \max_{1 \le i \le 64} |x_i|$ is used to align the spikes to a temporal reference. Aligning to the maximum absolute value is polarity-agnostic and thus robust to electrodes that record either negative-going or positive-going spikes. Following alignment, each waveform is cropped to a 48-sample window centred on the aligned peak as shown in Fig. 3(c). This window length is chosen to capture the full morphology of a typical extracellular action potential while excluding irrelevant baseline activity, thereby reducing input dimensionality and computational cost. A z-score normalization is then applied, which standardizes each waveform to have zero mean and unit variance:

$$x_i' = \frac{x_i - \mu}{\sigma}, \quad (1)$$

where $\mu$ and $\sigma$ denote the mean and standard deviation computed over all 48 samples of that waveform. This ensures a consistent feature distribution across different batches and epochs during training and calibration phases.

### B. Adaptive Level Crossing Algorithm

A key component of the proposed framework is the compression of spike waveforms into a compact yet informative representation. Rather than retaining all 48 samples of each aligned waveform, a nonuniform sampling strategy is adopted that selects $N = 8$ salient temporal points. This reduction provides a 6:1 compression ratio while preserving the discriminative morphology required for classification. A simple flowchart of the resulting adaptive level crossing algorithm is shown in Fig. 3(b). The proposed algorithm directly varies the sampling resolution by the slope ($S = y' = dy/dx$) and the curvature ($k = y''/(1 + (y')^2)^\wedge(3/2)$) profiles of the spikes.

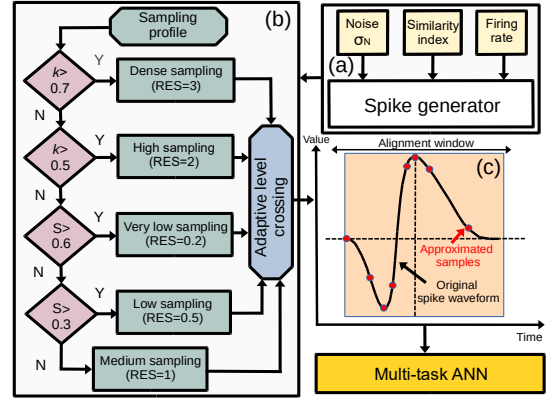

Fig. 3. (a) Spike generator, (b) flowchart of the adaptive level crossing sampling and (c) representative selected samples for a typical spike.

The proposed flowchart prioritizes curvature, using it as the primary indicator of spike waveform complexity. If the curvature exceeds $k > 0.7$, dense resolution sampling, i.e., RES=3, is utilized to preserve the spike samples. In regions of low curvature, the flowchart relies solely on a slope threshold to decide between "Low " and "Very Low" resolution, as gentle, flat terrain contains minimal detail and can be mapped very sparsely. The outcome of this cascading decision is a dynamic, non-uniform sampling profile as shown in Fig. 3(c). The system continuously cycles through this flowchart as it scans spikes, automatically adjusting its vertical step size ($\Delta V$) and point density in real-time.

### C. Multi-task Neural Network Architecture

The proposed ANN architecture, illustrated in Fig. 4, consists of an input layer, a shared feature extractor, and two task-specific branches. Each network input is a 48-sample spike waveform, obtained after the pre-processing pipeline of detection, alignment, cropping, and normalization. Formally, a single waveform is represented as a one dimensional tensor $x \in R^{48}$. During training, inputs are organized into mini batches of size $B$, resulting in a batch tensor of shape $[B, 48]$. Typical batch sizes range from 64 to 128, depending on hardware constraints and dataset size.

Formally, given an input waveform $x = \{x_1, x_2, \dots, x_{48}\} \in R^{48}$, the shared encoder parameterized by $\theta$ maps it into a $d$-dimensional latent representation $h = f_\theta(x) \in R^d$, where $f_\theta(\cdot)$ denotes the sequence of affine transformations and nonlinear activations implemented by the fully connected layers. This representation $h$ serves as the common feature space from which both compression and classification heads derive their outputs. The compression head is designed to predict the indices of the most informative high-curvature points that represent the compressed waveform. Specifically, the $d$-dimensional latent feature vector $h$ from the shared extractor is passed through a dedicated fully connected (FC) layer, producing an intermediate output of size $48 \times K$, where $K$ denotes the number of points to be selected. Each of the $K$ outputs corresponds to a categorical probability distribution over the 48 available time indices. Formally, the probability that the $j^{th}$ point occurs at index $i$ is:

$$p_{ij} = \frac{exp(z_{ij})}{\sum_{m=1}^{48} exp(z_{mj})}, i = 1,\ldots,48, j = 1,\ldots,K, \quad (2)$$

where $z_{ij}$ denotes the pre-softmax logit produced by the network. This formulation allows the compression head to express uncertainty about the optimal placement of informative points while being guided by the supervision provided during training. The final prediction is obtained by taking the maximum likelihood index for each of the $K$ distributions.

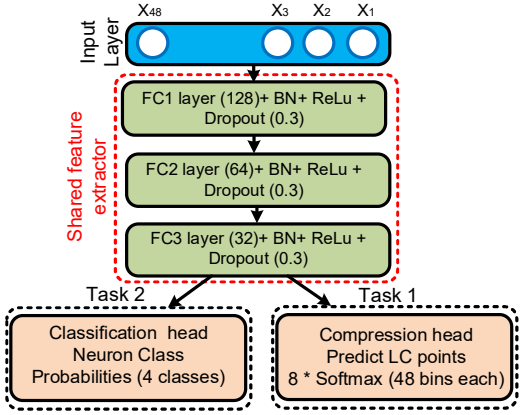

Fig. 4. Detailed architecture of the proposed multi-task ANN.

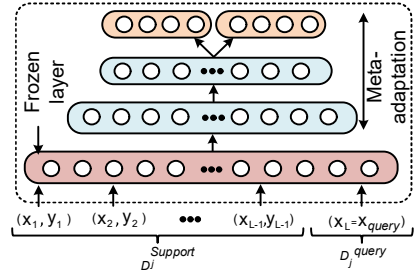

Fig. 5. Meta-transfer learning (MTL) to adapt the changes in recording channels. 4 way 4-shot [12] is used for fast adaptations. Support set has 4 classes ($N$=4) each class has 4 samples ($k$=4) with one query set ($q$=1).

Parallel to the compression pathway, the classification head maps the same shared representation $h$ into a categorical distribution over neuron classes and artefact. An additional FC layer followed by a softmax activation produces:

$$\hat{p}_c = \frac{exp(w_c^T h + b_c)}{\sum_{c'=1}^{C} exp(w_{c'}^T h + b_{c'})}, c = 1,\ldots,C, \quad (3)$$

where $C$ is the number of neurons, and $(w_c \ b_c)$ denote the trainable weights and biases for class $c$. The output vector $\hat{p} = [\hat{p}_1, \hat{p}_2, \ldots, \hat{p}_C]$ represents the posterior class probabilities for the input spike. The multi-task ANN adopts a categorical cross-entropy (CE) loss expressed as:

$$L_{Total} = L_{CE}^{class} + \alpha L_{CE}^{comp}, \quad (4)$$

where $L_{CE}^{class}$ denotes the cross-entropy loss for neuron class prediction, $L_{CE}^{comp}$ denotes the average cross-entropy loss across the $K = 8$ predicted compression points, and $\alpha$ is a weighting coefficient that balances the contribution of the two objectives.

### D. Meta-Transfer Learning (MTL)

The multi-task ANN jointly learns compression and classification from a fixed training set. However, in practical neural recording systems, distribution shifts occur frequently due to electrode drift, channel variability, or subject-specific differences [12]. Retraining of the full network would be computationally prohibitive and incompatible with real-time deployment on embedded devices. To address this challenge, MTL is proposed and integrated in the calibration process which enables rapid adaptation of the ANN using only a small support set of labelled spikes from the new recording channels. The shallow layers of the shared feature extractor which capture generic, low-level waveform features such as baseline fluctuations and global spike morphology are frozen during adaptation as shown in Fig. 5. Only the deeper layers of the network, which encode task-specific and channel-sensitive characteristics, are fine-tuned on the support set. This selective updating achieves two goals: it preserves stable features that generalize across channels, and it rapidly adjusts higher-level features to account for channel-specific variability.

Formally, let $\theta_s$ denote the parameters of the shallow layers and $\theta_d$ the parameters of the deeper layers. During adaptation, $\theta_s$ is kept fixed and $\theta_d$ is only updated using stochastic gradient descent on the joint loss $L_{Total}$ in Eq. (4). The parameter update rule for the deeper layers is:

$$\theta_d^{(t+1)} = \theta_d^{(t)} - \eta \nabla_{\theta_d} L(\theta_s, \theta_d^{(t)}; D_{Support}), \quad (5)$$

where $\eta$ is the learning rate and $D_{Support}$ denotes the small support set sampled from the new recording channel. This procedure prevents overfitting given the limited data size, while ensuring that the network adapts to channel-specific differences without retraining from scratch.

## III. TRAINING, IMPLEMENTATION AND EVALUATION SETUP

### A. Training of the Multi-task Neural Network

A dataset comprising more than 79,000 extracellularly recorded spike waveforms, also used in [1] and [3], was used to train the proposed multi-task ANN with MTL. Each spike is labelled based on ground truth into one of three neuron classes. The dataset was split into three disjoint subsets: 70% for training, 15% for validation, and 15% for final testing. Classes are distributed equally within all sets. The validation subset was used exclusively for hyperparameter tuning and early stopping, ensuring no test data leakage. This partitioning protocol provides a fair and reproducible basis for model evaluation.

Model training was conducted using mini-batch stochastic gradient descent with the Adam optimizer. Each mini-batch contained 128 spike waveforms. The initial learning rate was set to $1 \times 10^{-3}$ and reduced by a factor of 0.5 whenever the validation loss plateaued. To mitigate overfitting, batch normalization (BN) and a dropout rate of 0.3 was applied to all fully connected layers. Training proceeded for a maximum of 300 epochs, with early stopping triggered if the validation loss failed to improve for 20 consecutive epochs. The weighting factor in Eq. (4) was empirically set at $\alpha = 0.2$ to ensure compression fidelity. All experiments were conducted in MATLAB R2025a using the Deep Learning and Parallel Computing Toolboxes and accelerated on a workstation equipped with an NVIDIA RTX 3080 GPU and 32 GB of RAM.

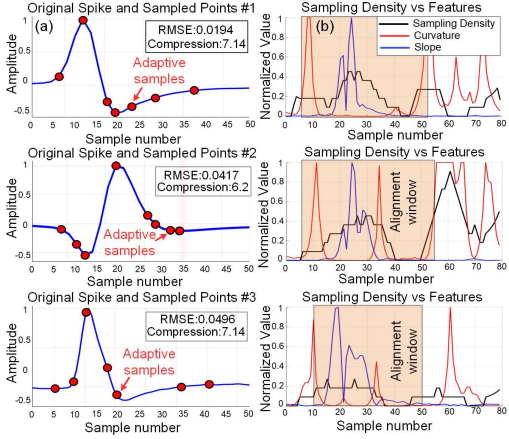

Fig. 6. (a) Demonstrating efficient compression (6×) while maintaining high fidelity (RMSE < 0.0194). (b) The method prioritizes sampling density in regions of high curvature.

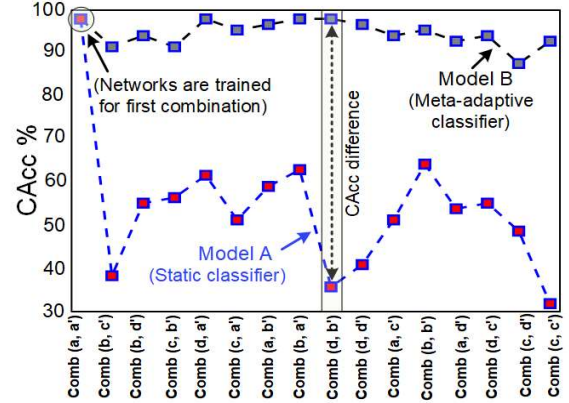

Fig. 7. Clustering accuracy (CAcc) comparison between the model A (static classifier) and model B (meta-adaptive) spike processors. CAcc is the classification accuracy (i.e. number of truly assigned feature vectors over the total number of feature vectors).

## B. Experimental Results and Analysis

To verify the efficacy and evaluate the performance of the proposed spike compression technique, a dataset composed of 500 different real spikes acquired from in-vivo, extra-cellular recordings was utilized. Fig. 6 shows the performance of the adaptive level crossing results for three spike waveforms. The graph in Fig. 6(a) superimposes the original waveforms and the approximated samples for all the 3 spikes. Averaged on all 500 spikes in the dataset, the overall sample-to-sample RMSE calculated based on normalized and reconstructed waveforms is 0.05. Fig. 6(b) shows the sampling density, curvature and the slope patterns of each spike shape. A dynamic test to evaluate the adaptivity of the processor was developed. To simulate dynamic variations in the data over time, a random data selection procedure was used. The neural simulator employed 4 standard combinations (a: Comb1, b: Comb2, c: Comb3, d: Comb4) with 3 spike shapes with different similarity indices and 4 noise standard deviations $\sigma_N$ (a': 0.05, b': 0.1, c': 0.015, d': 0.2) - i.e. 16 different combinations. Fig. 7 shows the classification accuracy. In model A (static classifier), the spike processor was configured to operate without MTL. In model B (meta-adaptive classifier), the multi-task ANN operates adaptively with the MTL. The CAcc comparison in Fig. 7 demonstrates the clustering performance superiority of the meta-adaptive classifier under variable input signal conditions. Model B utilizes a 4 way 4-shot for fast adaptations. This refers to the scenario where a model must rapidly adapt to distinguish between 4 entirely new classes (4-way) using only 4 labelled examples per category (4-shot). During each episode, the multi-task ANN receives a support set of 16 total examples (4 classes × 4 examples) to quickly adapt its knowledge, then is immediately tested on a query set from the same classes. Averaging the results in Fig. 7, yields an 63.12% median clustering accuracy for Model A and 94.43% for Model B.

## IV. CONCLUSION

A multitask ANN for efficient spike compression and reliable spike classification is introduced in this paper. The proposed framework integrates three core components: adaptive level crossing, a multi-task ANN, and MTL for rapid adaption to channel changes. Together, these components enable compact and discriminative representations of spike waveforms, reducing redundancy without sacrificing accuracy. In the future, the subsequent phase will focus on adapting the framework to accommodate datasets that contain a greater number of neuron classes, as observed in high-density probes.